\documentclass{article}[12pt]
\usepackage{graphicx}

\title{
Modelling light curves of binary systems: accounting for extended winds
}

\author{
E.A.ANTOKHINA, I.I.ANTOKHIN, \\  A.M.CHEREPASHCHUK
\\ Sternberg State Astronomical Institute, \\ Moscow
State University, Russian Federation\\ E-mail: elant@sai.msu.ru}

\begin{document}

\date{}
\maketitle

\abstract{
We suggest a simple synthesis model of an eclipsing binary system which includes one component with strong stellar wind. Numerical simulations show that the shape of the light curve (and in particularly the widths of  the minima) strongly depends on wind parameters. Wind effects are crucial in modelling light curves of binaries including e.g., WR stars.
}

\section{Introduction}

Several very massive Wolf-Rayet (WR) stars were recently discovered in eclipsing binaries, e.g. WR20a with WR mass \mbox{$M_{\rm WR}\approx 80\,M_\odot$} (Bonanas et al., 2004), NGC\,3601-A1 with \mbox{$M_{\rm WR1}\approx 116\,M_\odot$}, \mbox{$M_{\rm WR2}\approx 89\,M_\odot$} (Schnurr et al., 2008), etc. The radii of WR stars in these systems turned out to be surprisingly large ($\sim 20\,R_\odot$) (Bonanas et al., 2004; Moffat et al., 2004), similar to radii of main sequence (MS) stars of comparable masses. According to common evolutionary scenarios, a WR star is formed through a massive loss of its hydrogen-rich outer layers. One could expect that resulting stellar radii should be smaller than radii of MS stars with comparable masses.

The radii of the WR stars in the above binaries were obtained by analysing eclipses. Absorption in extended winds of WR stars was not accouned for, instead a standard Roche model (Wilson, 1979) of stars with thin atmospheres was used for both components of a binary. Including extended wind in a model is expected to reduce the radii and hopefully make them compatible with predictions of theory. Earlier attempts to account for wind absorption were made in  Pustyl'nik \& Einasto (1984), Antokhina \& Cherepashchuk (1988), Kallrath \& Milone (1999). In these papers, it was assumed that wind expands with constant velocity. The only advantage of such an assumption is simplicity of calculations. At the same time it is clearly inadequate to actual situation in hot star winds expanding with acceleration. In current paper, we present a simple wind model which allows one to use any parametric form of the velocity law for the wind.

\section{Model}

Roche model (Antokhina, 1988; Antokhina et al., 2000) is used for both components. One component is surrounded by homogeneous, spherically symmetric wind. While computing the model light curve, emission from every elementary area of each Roche surface is summed up, accounting for eclipses by stellar bodies and for absorption in the wind.  Velocity law is set by a parametric expression, e.g. the $\beta$-law: $v(r)=v_\infty\left(1-\frac{r_*}{r}\right)^\beta$.
The main opacity in optical continuum is electron scattering, optical depth from the point ($p,z_0$) in the wind (see Fig.1) is

\begin{equation}
\tau(p,z_0) = \int\limits_{z_0}^\infty \epsilon(z)\mathrm{d}z
\label{eq1}                                        
\end{equation} 

\noindent where the opacity coefficient 

\begin{equation}
 \epsilon(r)=\sigma_Tn_e(r)=\frac{z_e\sigma_T\dot{M}}{4{\pi}m_pv(r)r^2},
\end{equation}

\noindent $z_e$ is the number of electrons per unit atomic mass. If the chemical abundance of the wind is solar, hydrogen being the most abundant element, $z_e=1$ (in hot star winds hydrogen is ionized). In WR winds the most abundant element is helium, so $z_e \simeq 0.5$ if helium is fully ionized, $z_e \simeq 0.25$ in the He\,II zone. 

The model parameters along with their description are listed in Table~1.

\begin{figure*}[h!]
\begin{center}  
\includegraphics[width=5.0cm]{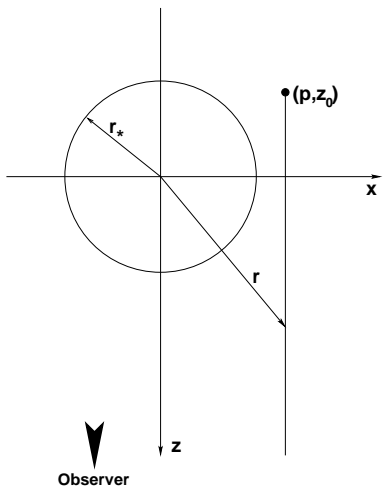}
\caption{The coordinate system for the wind model}
\end{center}
\end{figure*}

\newpage

{\footnotesize
{\bf Table 1. Input parameters of the model.}\\
 \begin{tabular}{ll}
 \hline\hline
 \noalign{\smallskip}
 Parameters & \hspace{1.5cm} Description  \\[+1mm]
 \hline\\
 \multicolumn{2}{c}{\bf Binary parameters}\\[+1mm]
 $M_1, M_2$ ($M_\odot$)     & Stellar masses \\
 $P$ (d)                    & Period \\
 $e$                        & Eccentricity \\
 $\omega$                   & Longitude of periastron, star No.1 \\
 $i$ (deg)                  & Orbital inclination \\
 $\mu_1$, $\mu_2$           & Roche lobe filling coefficients ($\mu=1$ for star filling \\
                            & its inner critical Roche lobe at periastron position) \\
 $T_1$,$T_2$ (K)            & Average effective temperatures of the components \\
 $\beta_1$,$\beta_2$        & Gravity darkening coefficients \\
 $A_1$,$A_2$                & Bolometric albedos \\
 $F_1, F_2$                 & Ratio of surface rotation rate to synchronous rate \\
 $x_1, x_2$                 & Limb darkening coefficients \\
 $l_3$                      & Third light \\
 $\lambda(n)$               & Effective wavelengths of monochromatic light curves\\[+1mm]
 \multicolumn{2}{c}{\bf Wind parameters}\\[+1mm]
 $\dot{M}$ ($M_\odot$/year) & Mass loss rate  \\
 $V_\infty$ (km/sec)        & Terminal velocity of the wind \\
 $\beta$                    & Parameter of the velocity $\beta$-law \\
 $z_e$                      & Number of electrons per unit atomic mass \\
 \noalign{\smallskip}
 \hline
 \end{tabular}
}

\section{Results}

To illustrate the influence of wind opacity on model light curves, we computed two models, a detached binary (Model 1) and a contact binary (Model 2). In each model stellar parameters of both components are identical but the secondary component is surrounded by a wind. This configuration helps to isolate the effects of the wind opacity on the light curve. When the wind is absent, the two minima of the light curve are identical. When the wind is ``turned on'', the depth of the secondary minimum is dramatically increased, while the depth of the primary minimum is decreased (in the secondary minimum the component surrounded by the wind is in front). The stellar parameters in both models were fixed while wind parameters varied. The values of model parameters are listed in Table 2. Wind parameters were varied one by one. While varying one parameter, the other two were fixed. The fixed values are $\dot{M}= 2 \cdot 10^{-5}\,M_\odot$/year, $V_\infty=2000$\,km/s, $\beta=0.5$. We assumed that wind consists of fully ionized helium ($z_e=0.5$).

\begin{figure*}
\begin{center}  
\includegraphics[width=4.5cm]{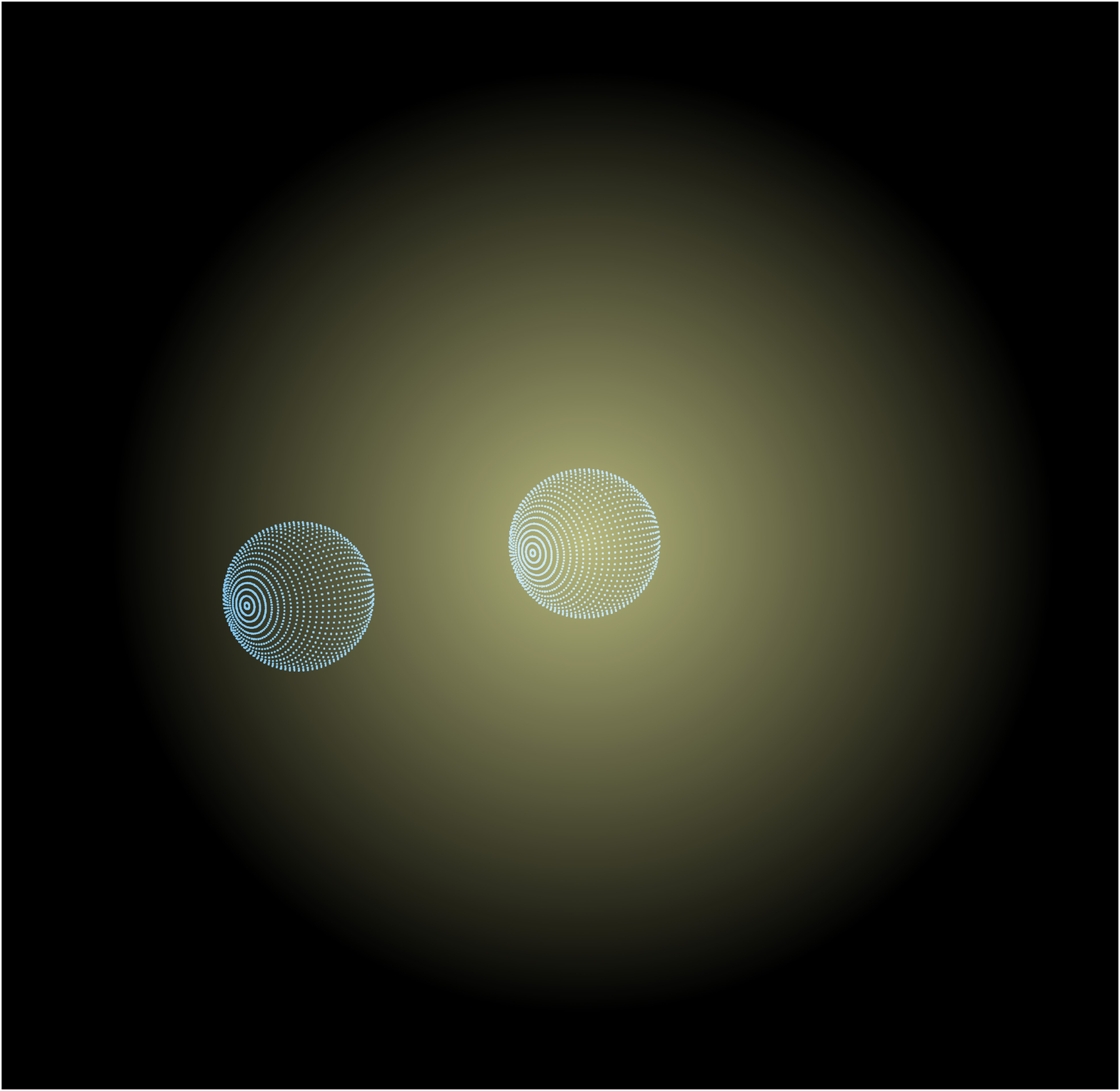}
\hspace{0.4cm}
\includegraphics[width=4.5cm]{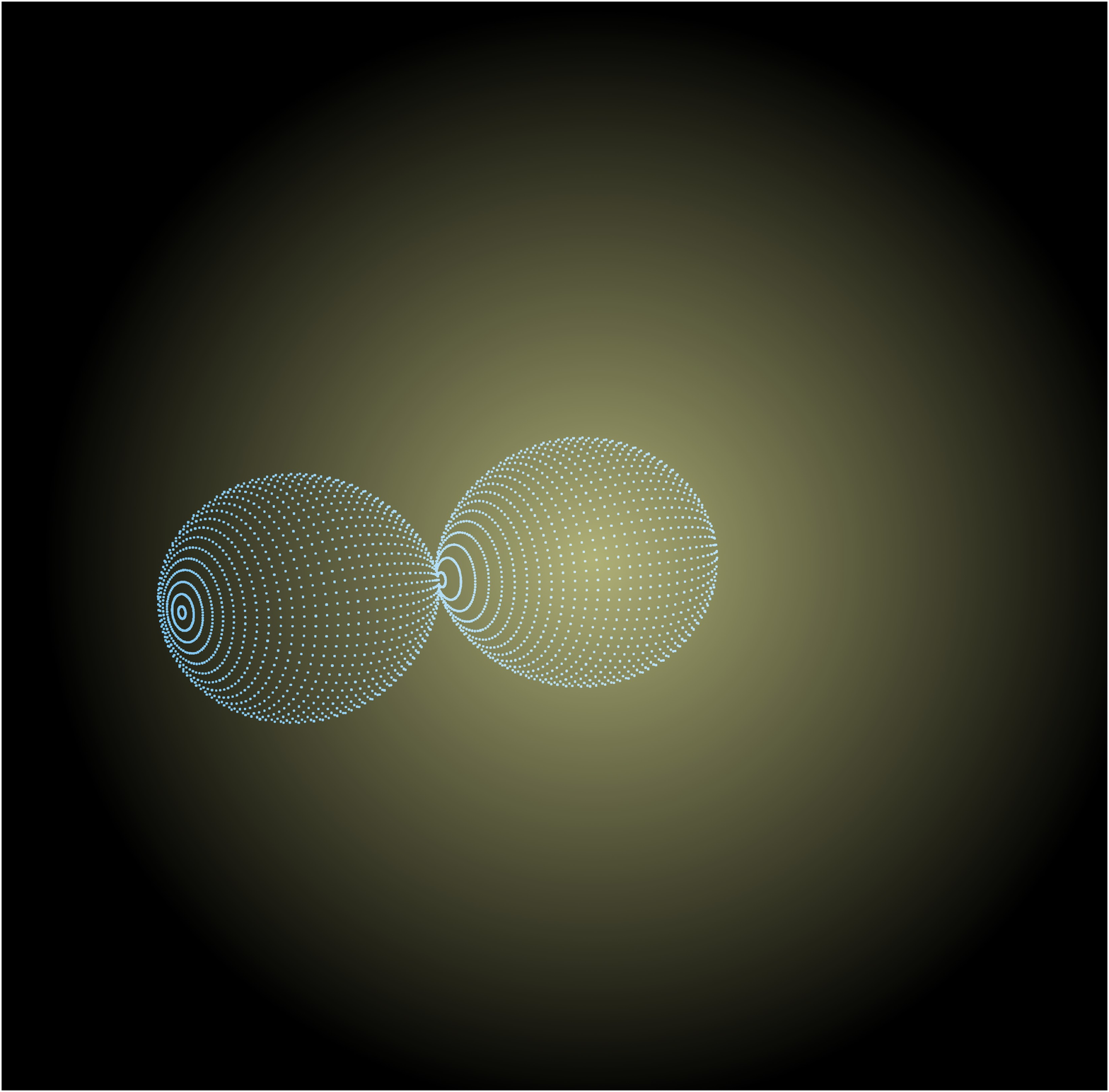}
\caption{Sky-plane view of Model 1 (left) and Model 2 (right)}
\end{center}
\end{figure*}

Simulations show that the shape of model light curves strongly depends on wind parameters (see Figs.3, 5). Clearly there is some degeneracy between the parameters. It is unlikely that the three most eccential wind parameters can be obtained independently from light curve analysis alone. Luckily, when solving light curves, most of them can be fixed at values obtained from independent sources (e.g., $V_\infty$ from UV spectra, $\dot{M}$ from radio data etc.). In light curve analysis, the increase in the width of the minima most importantly influences radii and temperatures of stars. Our simulations show that it is absolutely crucial to take into account extended winds while modelling light curves of binary systems including WR stars. 


This work was supported by the grant No 11-02-00258 from the Russian Foundation for Basic Research, the grant Nsh-7179.2010.2 from the Program for State Support of Leading Scientific Schools of Russian Federation, and by the grant RNP 2.1.1/12706.

\bigskip

{\footnotesize
{\bf Table 2. Parameters of simulated models.}\\[+1mm]
\begin{tabular}{lcc}
\hline\hline
\noalign{\smallskip}
Parameter & Model 1 (detached      & Model 2 (contact  \\
          & configuration) & configuration) \\
\noalign{\smallskip}
\hline\\
\multicolumn{3}{c}{\bf \hspace{2.2cm} Binary parameters}\\[+2mm]
$M_1$ ($M_\odot$)   &  $10$ & $10$ \\
$M_2$ ($M_\odot$)   &  $10$ & $10$ \\
$P$ (d)             & $5$  &  $5$ \\
$i$ (deg)           &  $80$  & $70$  \\
$\mu_1$             &  $0.5$  & $1.0$  \\
$\mu_2$             & $0.5$ & $1.0$   \\
$T_1 (K)$           & $30000$ & $30000$  \\
$T_2 (K)$           & $30000$ & $30000$  \\
$F_1,F_2$           & 1.0     & 1.0   \\
$\beta_1, \beta_2$  & 0.25    & 0.25  \\
$A_1,A_2$           & 1.0     & 1.0  \\
$x_1, x_2$          & 1.0  &  1.0   \\
$e$                 & 0. & 0. \\
$\omega$            & 0. & 0. \\
$l_3$               & 0.  & 0. \\
$\lambda$ (\AA)     & $4400$ & $4400$ \\[+1mm]
\multicolumn{3}{c}{\bf Wind parameters}\\[+1mm]
$\dot{M}$ ($M_\odot$/year) & $10^{-5}-4 \cdot 10^{-5}$ & $10^{-5}-4 \cdot 10^{-5}$ \\
$V_\infty$ (km/sec)        & $1000-2000$ & $1000-2000$ \\
$\beta$                    & $0.5-1.5$ & $0.5-1.5$ \\
$z_e$                        & $0.5$ & $0.5$ \\
\hline
\end{tabular}
}

\newpage 

\begin{figure*}[h!]
\begin{center}  
\includegraphics[width=5.0cm]{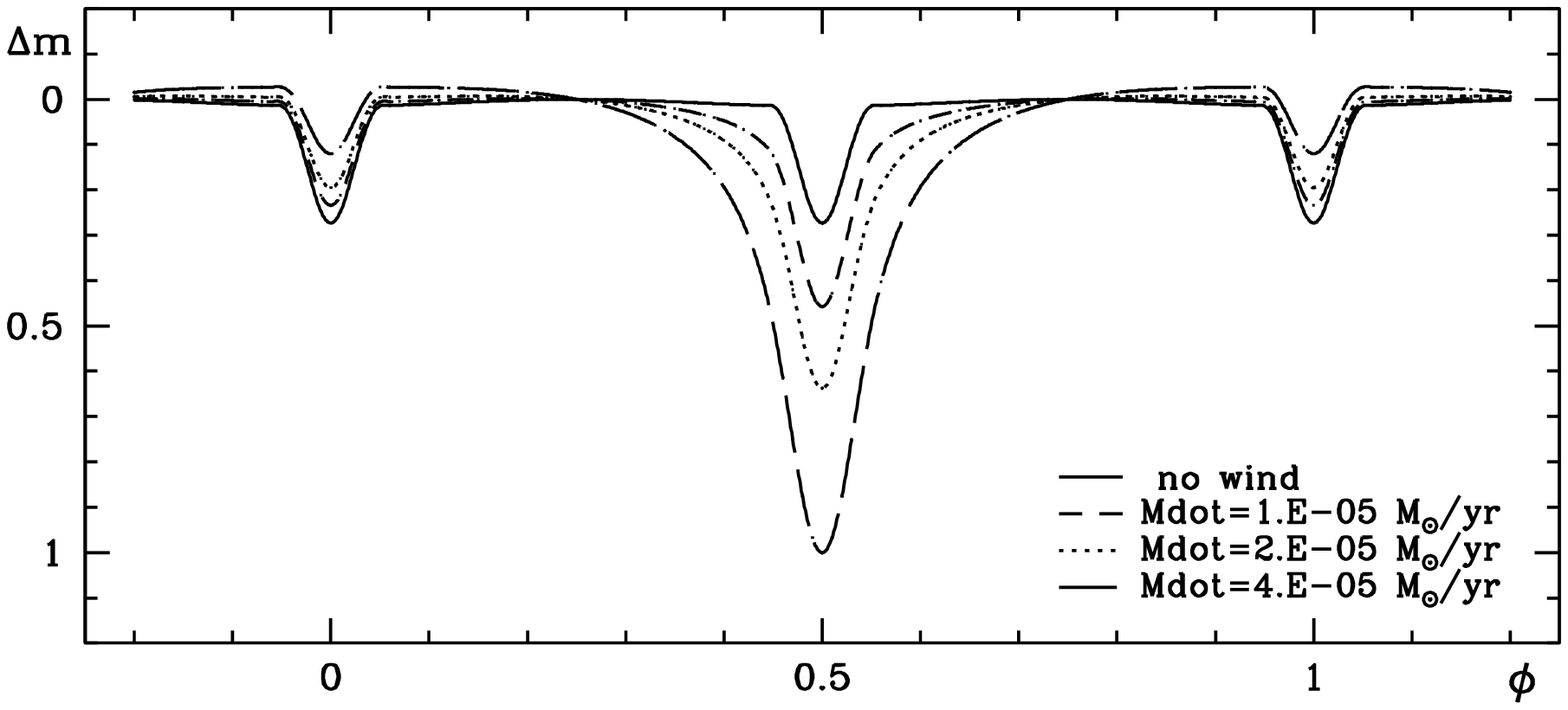}
\hspace{0.4cm}
\includegraphics[width=5.0cm]{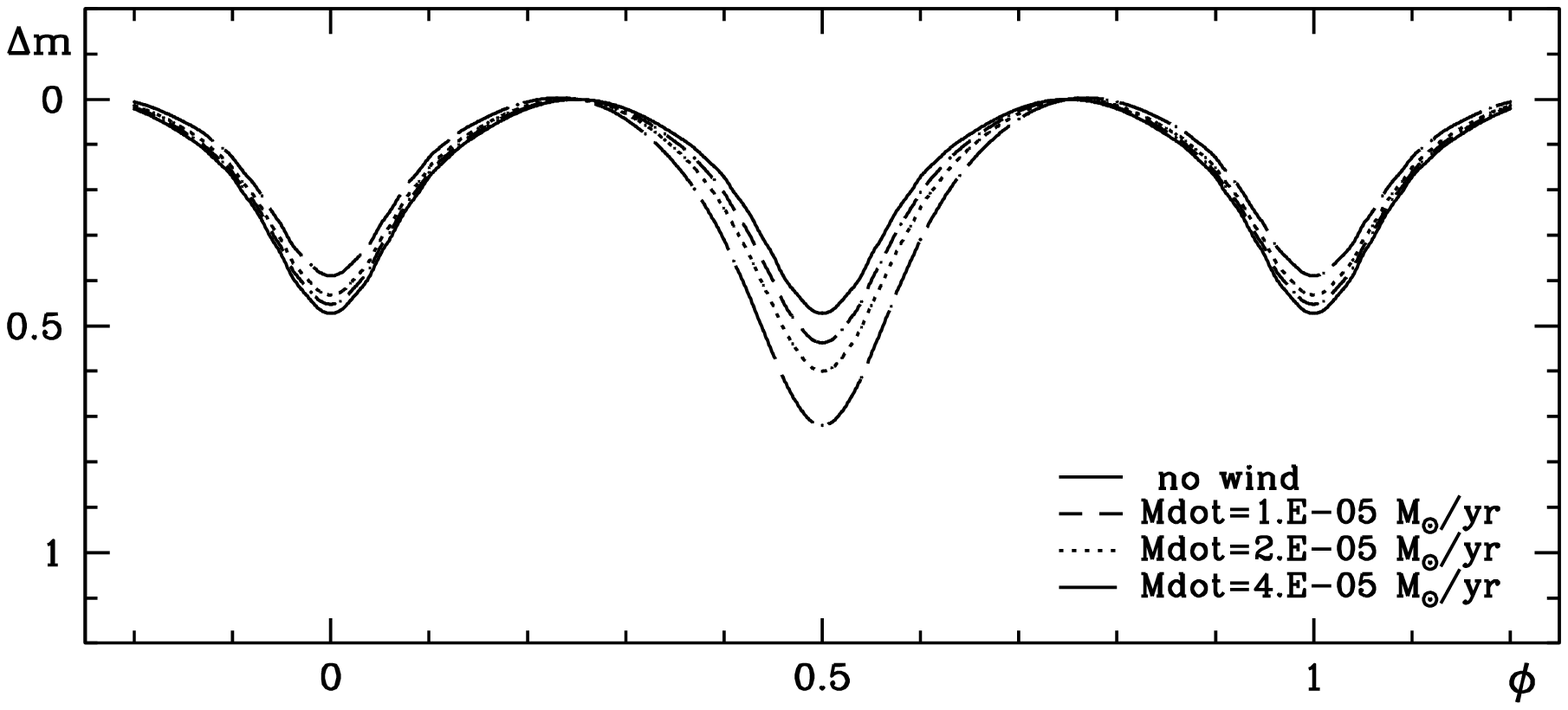}
\caption{Influence of mass loss rate $\dot{M}$ on the shape of light curve. Model 1 (left), Model 2 (right).}
\end{center}
\end{figure*}

\begin{figure*}[h!]
\begin{center}  
\includegraphics[width=5.0cm]{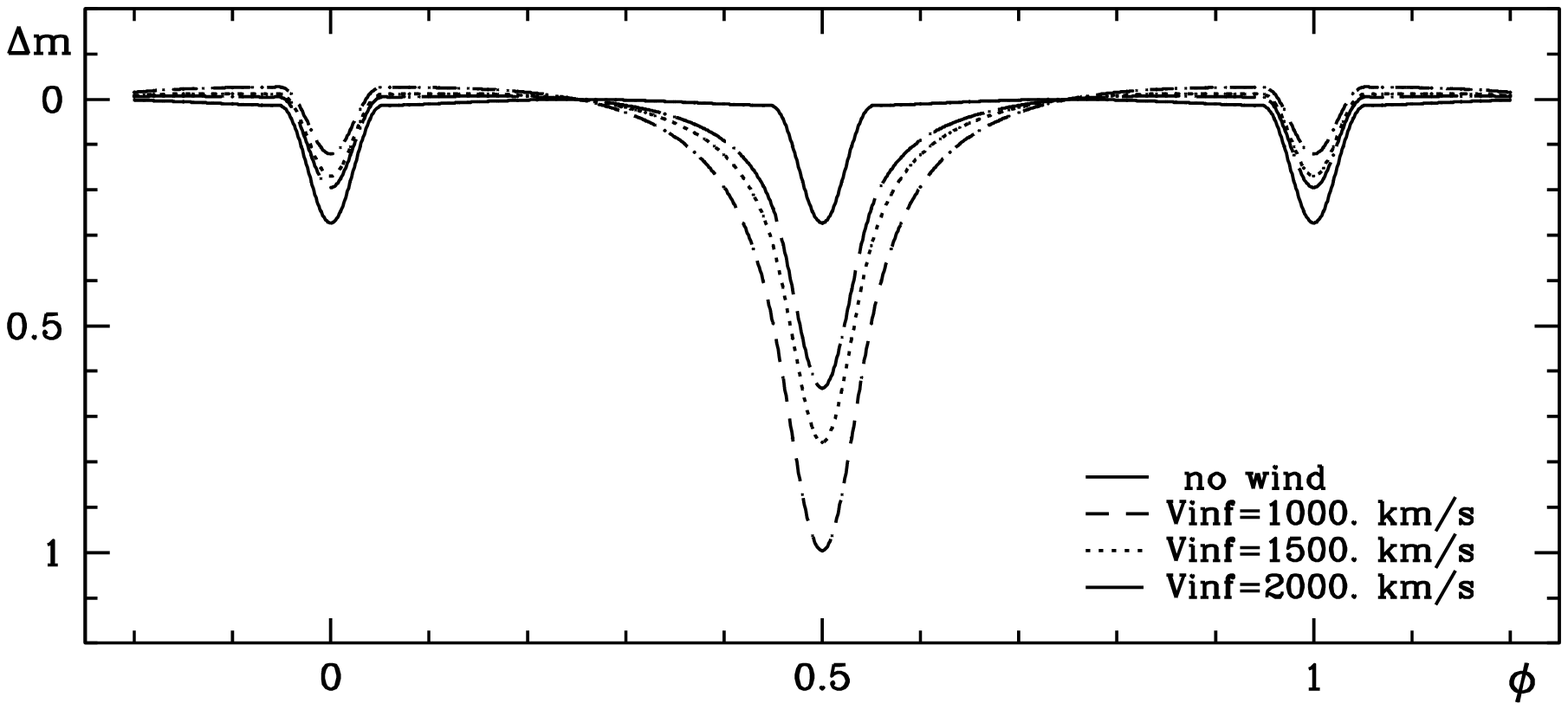}
\hspace{0.4cm}
\includegraphics[width=5.0cm]{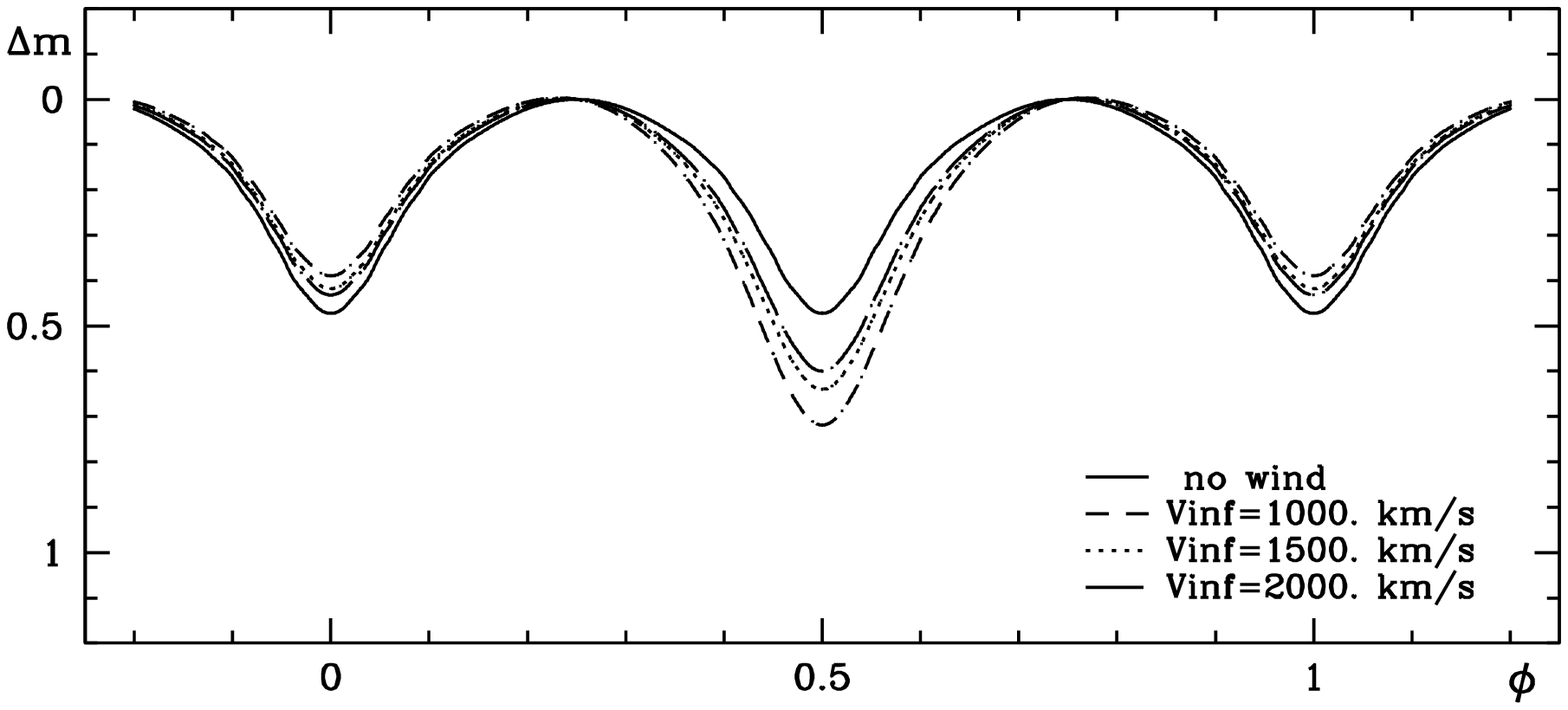}
\caption{Same as in Fig.3 , for the terminal wind velocity $V_\infty$}
\end{center}
\end{figure*}

\begin{figure*}[h!]
\begin{center}  
\includegraphics[width=5.0cm]{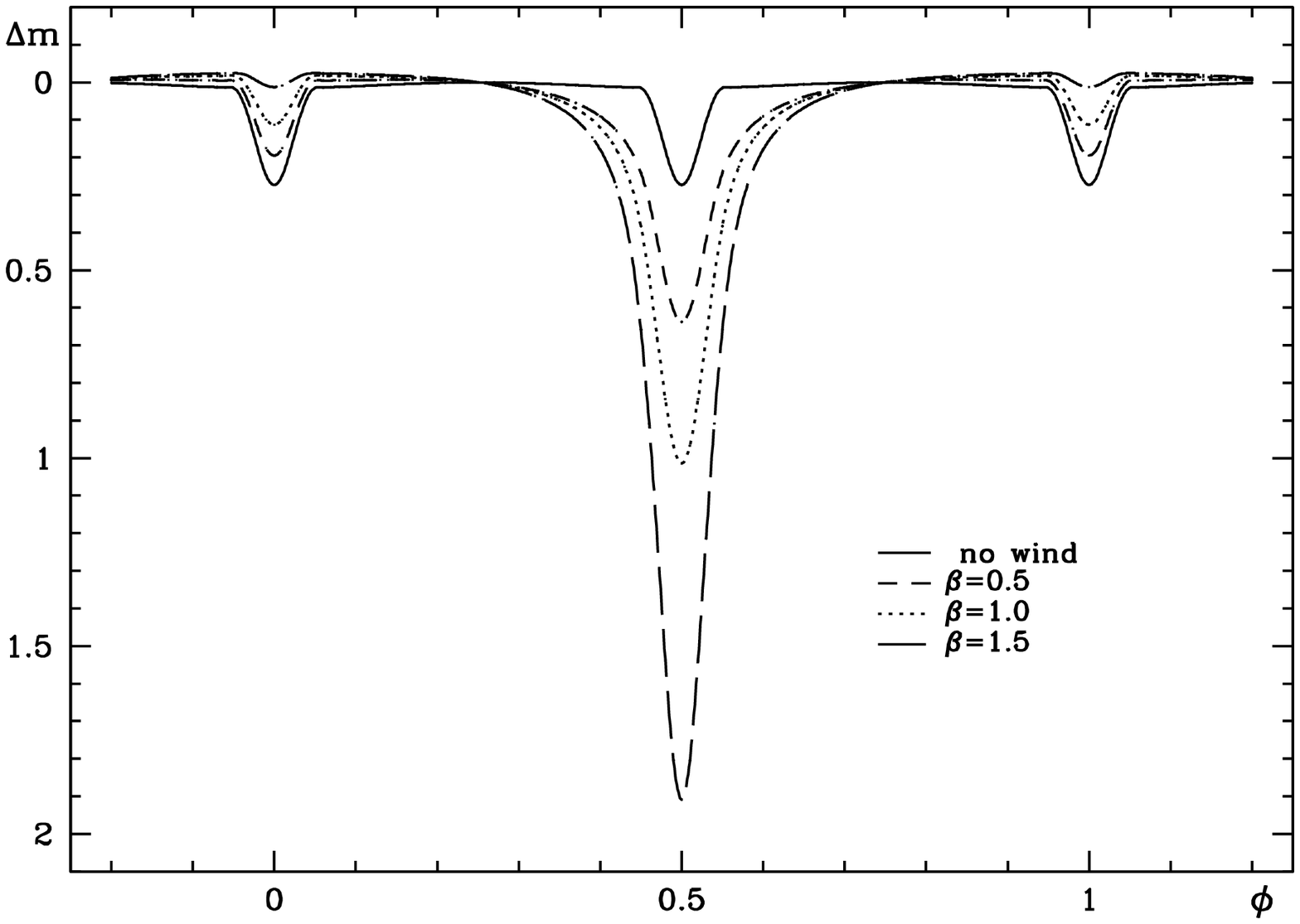}
\hspace{0.4cm}
\includegraphics[width=5.0cm]{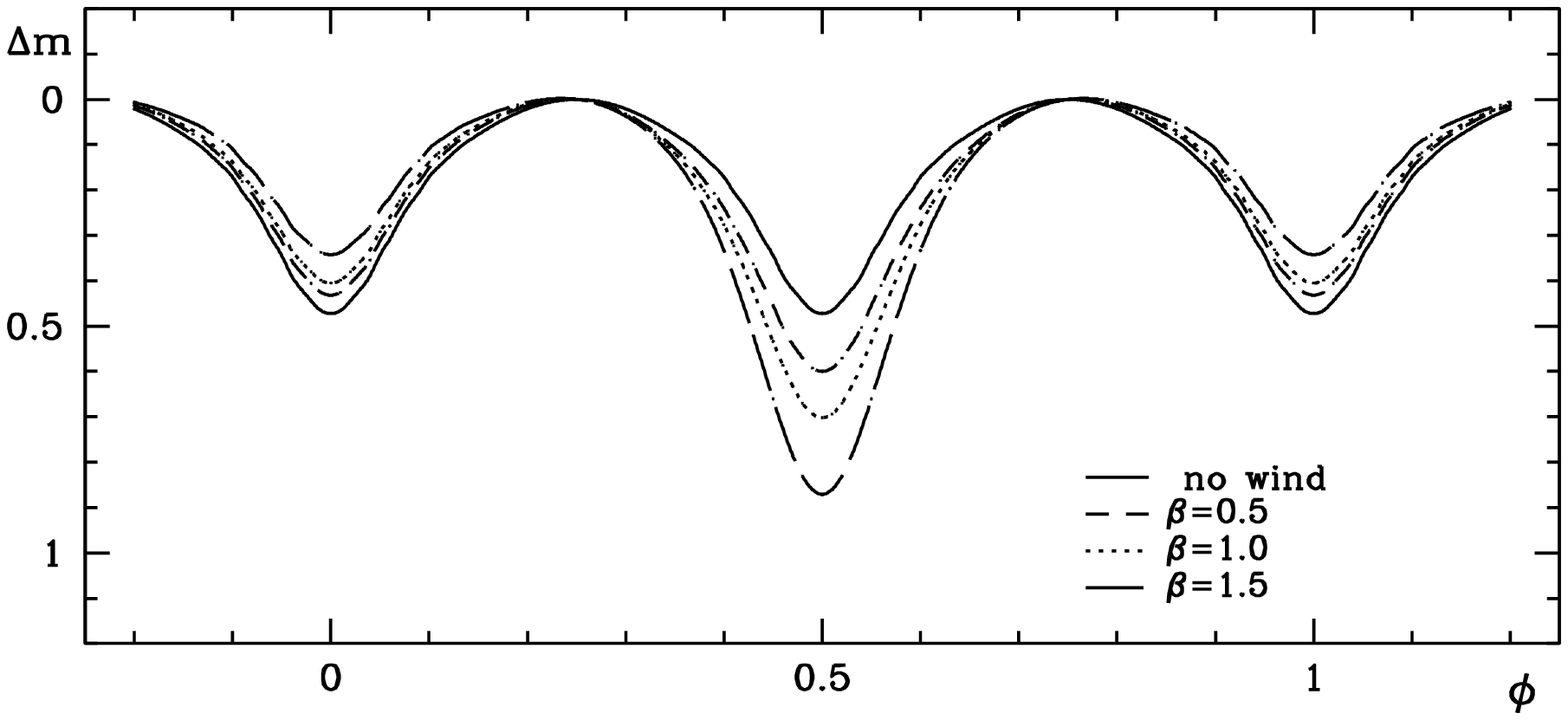}
\caption{Same as in Fig.3, for the parameter $\beta$ of the velocity law}
\end{center}
\end{figure*}


\noindent
{\bf References}


\noindent
Antokhina E.A., 1988, Sov. Astron., 32, 608

\noindent
Antokhina E.A., Cherepashchuk A.M., Sov. Astron., 1988, 65, 531

\noindent
Antokhina E.A., Moffat A.F.J., Antokhin I.I., et al., 2000, ApJ, 529, 463

\noindent
Bonanos A.Z., Stanek K.Z., Udalski A., et al., 2004, ApJ, 611, L33

\noindent
Kallrath J., Milone E., 1999, Eclipsing Binary Stars: Modeling and
Analysys, Springer-Verlag New York, 112

\noindent
Moffat A.F.J., Poitras V., Marchenko S.V., et al., 2004,
Astron. J., 2004, 128, 2854

\noindent
Pustyl'nik I.B., Einasto L., 1984, Sov. Astron. Lett., 10, 215

\noindent
Schnurr O., Casoli J., Chene A.-N., et al., 2008,  MNRAS, 389, L38

\noindent
Wilson, R.E. 1979, ApJ, 234, 1054 


\end{document}